\documentclass[aps,prc,showpacs,superscriptaddress,preprintnumbers,floatfix,nofootinbib]{revtex4}
\usepackage{multirow}
\usepackage{hhline}
\usepackage{bbm}
\usepackage{epsfig}
\usepackage{graphics}
\usepackage{graphicx}
\usepackage{dcolumn}
\usepackage{bm}
\usepackage{float}
\usepackage{amssymb}
\usepackage{amsmath}
\usepackage{amsfonts}
\usepackage[percent]{overpic}
\usepackage{natbib}
\usepackage[colorlinks,linkcolor=blue,urlcolor=blue,citecolor=blue]{hyperref}
\usepackage[usenames]{color}
\usepackage{CJK}

\begin{document}
\begin{CJK*}{GB}{gbsn}
\title{Dipole oscillation modes in light $\alpha$-clustering nuclei}

\author{W. B. He}
\affiliation{Shanghai Institute of Applied Physics, Chinese Academy of Sciences, Shanghai 201800, China}
\affiliation{University of the Chinese Academy of Sciences, Beijing 100080, China}

\author{Y. G. Ma \footnote{Email: ygma@sinap.ac.cn}}
\affiliation{Shanghai Institute of Applied Physics, Chinese Academy of Sciences, Shanghai 201800, China}
\affiliation{ShanghaiTech University, Shanghai 200031, China}

\author{X. G. Cao \footnote{Email: caoxiguang@sinap.ac.cn}}
\affiliation{Shanghai Institute of Applied Physics, Chinese Academy of Sciences, Shanghai 201800, China}

\author{X. Z. Cai }
\affiliation{Shanghai Institute of Applied Physics, Chinese Academy of Sciences, Shanghai 201800, China}
\author{G. Q. Zhang}
\affiliation{Shanghai Institute of Applied Physics, Chinese Academy of Sciences, Shanghai 201800, China}

\date{\today}
\begin{abstract}
The $\alpha$ cluster states are discussed in a model frame of extended quantum molecular dynamics.
Different alpha cluster structures are studied in details, such as $^8$Be two-$\alpha$ cluster structure, $^{12}$C triangle structure, $^{12}$C chain structure, $^{16}$O chain structure, $^{16}$O kite structure, and $^{16}$O square structure.
The properties studied, include as the width of wave packets for different $\alpha$ clusters, momentum distribution, and the binding energy among $\alpha$ clusters.
It is also discussed  how the $\alpha$ cluster degree of freedom affects nuclear collective vibrations.
The cluster configurations in $^{12}$C and $^{16}$O are found to have corresponding characteristic spectra of giant dipole resonance (GDR), and the coherences of different $\alpha$ clusters's dipole oscillation are described in details.
The geometrical and dynamical symmetries of $\alpha$-clustering configurations are responsible for the number and centroid energies of peaks of  GDR spectra. Therefore, the GDR can be regarded as an effective probe to diagnose different $\alpha$ cluster configurations in light nuclei.
\end{abstract}
\pacs{21.60.Gx, 24.10.-i, 24.30.Cz, 25.20.-x}
\maketitle
\section{\label{sec:1}Introduction}

Clustering is a fundamental physics aspects in light nuclei lighter$\left(Z\leq16\right)$, where the mean filed effect is not strong enough to break cluster formation at low temperatures.
It is typically observed as excited states of those nuclei and also in the ground states for nuclei far from the $\beta$ stability line, where nuclei can  behave like molecules composed of nucleonic clusters.
Many authors have focused on clustering over the past decades \cite{greiner1995nuclear,Vonoertzen2006PR432}.
Near the threshold of decay into the subunit, nuclei can be assumed to change into the molecule-like structures \cite{Ikeda1968PTPSE68}.
Due to high stability of the $\alpha$ particle, the 2n-2p correlation plays a critical role in light nuclei clustering.
The self-conjugate light nuclei are expected to have a phase change and nucleons condense into $\alpha$-particles, as the density is lower than one third of the normal nuclear matter density \cite{Brink1973npa}.
As the density falls to one fifth of the normal nuclear matter, the self-conjugate light nuclei are expected to be in an $\alpha$-gas or a Bose condensed state \cite{schuck2003alpha}.
In neutron rich light nuclei, nuclear molecules with clusters bound via neutrons can show up, at low density \cite{Vonoertzen2006PR432}.
As the density decreases, $\alpha$ clustering will dramatically change the nuclear equation of state \cite{Girod2013PRL111,Ropke2013NPA897,Horowitz2006NPA776,Natowitz2010PRL104,Qin2012PRL108}.
The famous Hoyle state in $^{12}$C at 7.65 MeV, which is considered as a key point of the $^{12}$C synthesis in the Universe, is believed to be formed out of a weakly interacting gas of $\alpha$ particles \cite{Holy}.
However, many issues have not yet been well understood, such as how $\alpha$ clustering determines the configurations and shapes of the many-body system, what aspects, and the underlying mechanism, are collective dynamics of $\alpha$ clustering systems, {\it etc}. \cite{Umar2010PRL104,Ichikawa2011PRL107,Ebran2012Nature487,FREER2007RPP70,FUNAKI2010JPG37}.

Since its discovery, giant dipole resonances (GDR) has been revealed in the nuclei as light as $^4$He \cite{Efros1997PRL} and as heavy as $^{232}$Th \cite{Berman1975RMP}.
Therefore, GDR is a good tool for systematical investigation on collective properties throughout the nuclear chart.
As the most pronounced feature for excited nuclei, GDR can give crucial clues to understand nuclear structure and collective dynamics.
The centroid energy of GDR can provide direct information about nuclear size and the nuclear equation of state \cite{Harakeh2001giant}.
Meanwhile, the GDR width can be used as a direct experimental probe to measure the nuclear deformation at finite temperature and angular momentum over the entire mass region \cite{Pandit2010PRC81,Pandit2013PRC87}.
The GDR strength has a single peak distribution for spherical nuclei with mass number $>$ 60. However,  the GDR strength usually shows configuration splitting in light nuclei
\cite{Eramzhyan1986PR136,Harakeh2001giant,Yamagata2004PRC69,He2014PRL113}.
In light nuclei with molecule-like structures, the deformation is huge enough to cause big splitting of GDR.
In addition, the degree of freedom of clusters in nuclei affects the GDR spectra.
Multifragmented peaks can be expected for self-conjugate ($\alpha$)-nuclei with a prominently developed $\alpha$ cluster structure in excited states.
A recent study by Y. Chiba, et al. found that asymmetric cluster configurations in $\alpha$ conjugate nuclei contribute to resonances by isoscalar dipole transition at relatively small excitation energy \cite{Chiba2016PRC93}.
Therefore, it is interesting to study how an $\alpha$ cluster component manifests itself in GDRs.
The GDR spectra shall provide important and direct information to reveal the geometrical configurations and dynamical interactions among $\alpha$ clusters.

Configurations of an $\alpha$ clusters is a key problem to understand the clustering in light nuclei.
Theoretical predictions made recently on $\alpha$ cluster configuration in light nuclei revealed the following aspects.
$^8$Be composed by two $\alpha$ particles, has a scarcely greater value than the threshold energy for the decay into two $\alpha$ particles\cite{Ikeda1968PTPSE68}.
For $^{12}$C, triangular-like configuration, is predicted around the ground state by Fermionic molecular dynamics \cite{Chernykh2007PRL98}, anti-symmetrized molecular dynamics \cite{Kanada-En'yo2012PTEP2012,Furuta2010PRC82}, and covariant density
functional theory \cite{Liu2012CPC36}, which is supported by a recent experimental result \cite{addref1}.
A three-$\alpha$  linear-chain configuration was predicted as an excited state in time-dependent Hartree-Fock theory \cite{Umar2010PRL104}, and other different approaches \cite{Morinaga1966PL21}.
In the framework of the cranking covariant density functional theory, the mechanisms to stablize a linear-chain configuration was discussed in detail \cite{Zhao2015PRL115}.
The intrinsic density of $^{12}$C and $^{16}$O may display localized linear-chain density profile as an excitation of the condensed gas-like states described with the Brink wave function and the Tohsaki-Horiuchi-Schuck-R\"{o}pke wave function(THSR) \cite{THSR2001PRL87,schuck2003alpha,Suhara2014PRL112}.
For $^{16}$O, the linear-chain structure with four-$\alpha$ clusters  was supported by the $\alpha$ cluster model \cite{Bauhoff1984PRC29} and the cranked Skyrme Hartree-Fock method \cite{Ichikawa2011PRL107}.
A tetrahedral structure of  $^{16}$O, made out of four-$\alpha$ clusters, is found above the ground state with the constrained Hartree-Fock-Bogoliubov approach \cite{Girod2013PRL111}.
However, recent calculations with nuclear chiral effective field theory \cite{Epelbaum2014PRL112} and covariant density functional theory \cite{Liu2012CPC36} support the tetrahedral $\alpha$ configuration located at the ground states.
An algebraic model \cite{Bijker2014PRL112} shows that the ground-state rotational band supporting the nucleus has tetrahedral symmetry.
Orthogonality condition model calculations show a duality of the mean-field-type as well as $\alpha$-clustering character in the $^{16}$O ground state \cite{Yamada2012PRC85}.
There are different configuration descriptions implying the $\alpha$ cluster structure in  $^{20}$Ne and $^{24}$Mg, such as three-dimensional shuttle shape \cite{Girod2013PRL111,Ebran2012Nature487} or chain states \cite{Chappell1995PRC51,Marsh1986PLB180} and non-localized cluster states \cite{ZHOU2013PRL110}.
Therefore, it is important to look for new experimental probes to diagnose different configurations for $\alpha$-conjugate nuclei around the cluster decay threshold \cite{Broniowski2014PRL112}.

In this work, we report our results of GDRs of $\alpha$ cluster states of $^8$Be, $^{12}$C, and $^{16}$O within a microscopic dynamical many-body approach.
First, we discuss the method of GDR calculations within QMD models.
Then by demonstrating the results of $^{12}$C and $^{16}$O in the ground states, we show the reliability of GDR calculations in our model, and propose the coexistence of triangle shape and spherical shape in $^{12}$C ground states.
Finally, we investigate how the different $\alpha$ configurations lead to multifragmented peaks of GDR and the underlying mechanism which is responsible for the collective motion of $\alpha$-clustering light nuclei reported in our previous publication \cite{He2014PRL113}.

\section{\label{sec:2}Model and methodology}
\subsection{\label{sec:2.1}Model introduction}

Quantum molecular dynamics (QMD), a powerful tool for studying intermediate energy nuclear reactions and nuclear fragmentation \cite{Aichelin1991PR}, has been successfully applied in studies of giant resonances of GDR, pygmy dipole resonance(PDR) and giant monopole resonance(GMR) due to its microscopic basis and high flexibility \cite{Kanada-En'yo2005PRC72,Furuta2010PRC82,Wu2010PRC81,Tao2013PRC87,CAO2010PRC81}.
In the following calculations of GDRs, the nuclear system is described within the QMD model framework.
To apply this approach to light nuclei like $^8$Be, $^{12}$C, and $^{16}$O, some requirement for the model are necessary.
For example, the energy, radius of ground states shall be well described, and the ground states shall be stable enough.
Nevertheless, standard QMD shows insufficient stability because the initialized nuclei are not in their real ground states.
In this paper, we use an extended QMD (EQMD) of some new features \cite{Maruyama1996PRC53,Wada1998PLB422}.
It is introduced briefly as follows.

In EQMD, nucleons are treated as Gaussian wave packets ${{\rm{\varphi }}_i}$, which are written as:
\begin{equation}
{{\rm{\varphi }}_i}\left( {{{\bf r}_i}} \right) = {\left( {\frac{{{v_i} + v_i^ * }}{{2\pi }}} \right)^{3/4}}\exp \left[ { - \frac{{{v_i}}}{2}{{\left( {{{\bf r}_i} - {{\bf R}_i}} \right)}^2} + \frac{i}{\hbar }{{\bf P}_i} \cdot {{\bf r}_i}} \right],
\end{equation}
where ${v_i}$ ${v_i} = {{1/{\lambda _i}}} + i{\delta _i}$ is width of the complex Gaussian wave packets.
$\lambda$ and $\delta$ are dynamic variables.
The ${v_i}$ of Gaussian wave packets for each nucleon is dynamic and independent.
This is an important improvement compared with the standard QMD, in which a uniform and static width is applied for all nucleons.
Dynamical wave-packet width not only improves the capability of describing ground state, but also helps to describe nuclear exotic structures, such as nuclear halo structure.
Furthermore, the kinetic-energy term arising from the momentum variance of wave packets is taken into account by subtracting the spurious zero-point center of mass (c.m.) kinetic energy from the Hamiltonian.
This procedure is important for QMD models to describe nuclear cluster states and fragmentation.
In standard QMD, the kinetic-energy term arising from the momentum variance of wave packets is constant.
Thus, the constituent nucleons having finite momenta are not in energy-minimum states, hence the source of insufficient stability.
So, the Hamiltonian can be written as
\begin{equation}
\begin{array}{l}
H = \left\langle {\bf \Psi}  \right|\sum\limits_i { - \frac{{{\hbar ^2}}}{{2m}}} \nabla _i^2 - {\widehat T_{c.m.}} + {\widehat H_{{\mathop{\rm int}} }}\left| {\bf \Psi}  \right\rangle \\
 = \sum\limits_i {\left[ {\frac{{{\bf P}_i^2}}{{2m}} + \frac{{3{\hbar ^2}\left( {1 + \lambda _i^2\delta _i^2} \right)}}{{4m{\lambda _i}}}} \right]}  - {T_{c.m.}} + {H_{{\mathop{\rm int}} }},
\end{array}
\end{equation}
where $T_{c.m.}$ is the term of zero-point center of mass (c.m.) kinetic energy, the form of which can be found in details in Ref. \cite{Maruyama1996PRC53}.
For the effective interaction, Skyrme and Coulomb forces, the symmetry energy, and the Pauli potential are used,
\begin{equation}
H_{int}=H_{Skyrme}+H_{Coulomb}+H_{Symmetry}+H_{Pauli}.
\end{equation}
The form of Skyrme interaction use in EQMD model is the simplest, written as
\begin{eqnarray}
H_{Skyrme}=\frac{\alpha }{2\rho _{0}}\int \rho^2\left ( \mathbf{r} \right )d^3 r+\frac{\beta }{\left ( \gamma +1 \right )\rho _{0 }^{\gamma}}\int \rho^{\gamma +1}\left ( \mathbf{r} \right )d^3 r,
\end{eqnarray}
where $\alpha$=-124.3 MeV, $\beta$=70.5 MeV, and $\gamma$=2.
The symmetry potential is written as
\begin{eqnarray}
H_{Symmetry}=\frac{C_{S}}{2\rho _{0}}\sum_{i,j\neq i} \int \left [ 2\delta \left ( T_i,T_j \right )-1 \right ]\rho_i\left ( \mathbf{r} \right )\rho_j\left ( \mathbf{r} \right )d^3r,
\end{eqnarray}
where $C_S$ is the symmetry energy coefficient and here $C_S$=25 MeV.
Specifically, the Pauli potential is written as
\begin{eqnarray}
H_{Pauli} = \frac{c_P}{2}\sum_i(f_i-f_0)^\mu\theta(f_i-f_0),\\
f_i\equiv\sum_j\delta(S_i,S_j)\delta(T_i,T_j)|\langle\phi_i|\phi_j\rangle|^2,
\end{eqnarray}
where, $f_i$ is the overlap of a nucleon $i$ with nucleons having the same spin and isospin; $\theta$ is the unit step function; $c_P$ is a coefficient related to strength of the Pauli potential.
This potential inhibits the system from collapsing into the Pauli-blocked state at low energy and gives the model capability to describe $\alpha$-clustering.
This capability is crucial to our calculation because it enable the GDR study on $\alpha$ cluster configurations.
Since the clustering configurations and the profiles of GDR spectra are not sensitive to different forms of potential,  the relation between clustering configurations and GDR spectra is independent of EQMD model.
The phase space of nucleons is obtained initially from a random configuration.
To get the energy-minimum state as a ground state, a frictional cooling method is used for the initialization process.
The model can describe quite well the ground state properties, such as binding energy, rms radius, deformation, {\it etc.}, over a wide mass range \cite{WangSS}.

\subsection{\label{sec:2.2}GDR algorithm}

The macroscopic description of GDR by the Goldhaber-Teller model \cite{Goldhaber1948PR74} assumes that protons and neutrons collectively oscillate with opposite phases in an excited nucleus.
In the EQMD model, the location and momentum of all nucleons are explicit variables.
Based on the Goldhaber-Teller assumption, we can calculate the oscillation energy spectra.
The dipole moments of the system in coordinate space $D_{G}(t)$ and momentum space $K_{G}(t)$ are, defined as follows \cite{Baran2001NPA679,Wu2010PRC81,Tao2013PRC87}:
\begin{eqnarray}
D_{G}(t) = \frac{NZ}A\bigg[R_Z(t)-R_N(t)\bigg],\\
K_{G}(t) = \frac{NZ}{A\hbar}\bigg[\frac{P_Z(t)}Z-\frac{P_N(t)}N\bigg],
\end{eqnarray}
where, $R_Z(t)$ and $R_N(t)$, and $P_Z(t)$ and $P_N(t)$, are the c.m.'s of the protons and neutrons in coordinate and momentum space, respectively.
$N$ is the neutron number; and $A$ is the mass number.
$K_{G}(t)$ is the canonically conjugate momentum of $D_{G}(t)$.
The evolution of the excited wave function to the final state is obtained by the EQMD model.
From the Fourier transform of the second derivative of $D_{G}(t)$ with respect to time, i.e.,
\begin{equation}
D^{''}(\omega) = \int_{t_0}^{t_{max}}D^{''}_G(t)e^{i\omega t}dt,
\label{eq:Fourier}
\end{equation}
the dipole resonance strength of the system at excitation energy $E = \hbar \omega$ can be obtained by Eq.\ref{eq:dpde},
\begin{equation}
\label{eq:dpde}
\frac{dP}{dE} = \frac{2e^2}{3\pi\hbar c^3E}\big|D^{''}(\omega)\big|^2,
\end{equation}
where dP/dE can be interpreted as the nuclear photo-absorption cross section.

\begin{figure}[H]
\centering
\includegraphics[width=.5\textwidth]{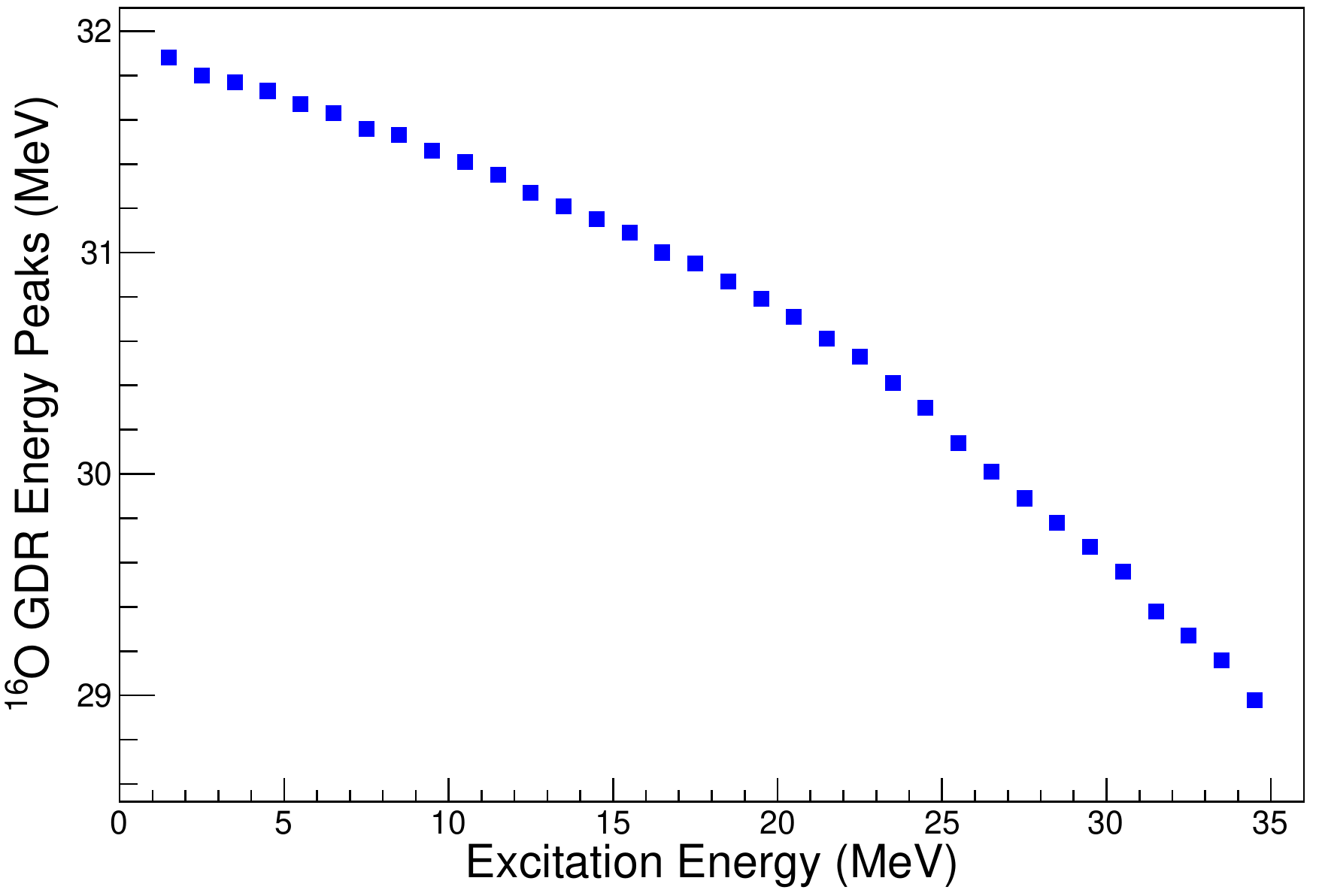}
\vspace{-0.3cm}
\caption{\footnotesize (Color online) Dependence of $^{16}$O GDR energy peaks on excitation energy.}
\label{fig:peak}
\end{figure}

Also, the GDR cross section can be obtained by calculating nuclear response to external excitation. To describe an excitation of external dipole field, the dipole operator can be written as
\begin{equation}
\mathbf{R}=\sum_{i}\left ( \frac{N}{A} \mathbf{Pr}_i-\frac{Z}{A} \mathbf{Ne}_i\right )\mathbf{r},
\end{equation}
where $i$ indexes all nucleons in the nucleus;
N is the neutrons number;
P is the protons number; and
A is the mass number.
${\bf Pr}$ and $\mathbf{Ne}$ are the projection operators for protons and neutrons, respectively.
The dipole excitation can be written as an additional perturbative component to the Hamiltonian,
\begin{equation}
H'=\mathbf{R}\varepsilon \delta \left ( t \right ),
\end{equation}
where $\varepsilon$ is an arbitrary small value, and $\delta P$ is the variation of momentum.
The system wave function can be written as
\begin{equation}
\begin{array}{l}
\left| {\Psi \left( {t = 0} \right)} \right\rangle  = \exp \left[ { - i\int {H'dt} } \right]\left| {\Psi \left( 0 \right)} \right\rangle \\
 = \exp \left[ { - i\frac{r}{\hbar } \cdot \sum\limits_i {\Delta P\left( {\sqrt {\frac{N}{{AZ}}} \widehat {{{\Pr }_i}} - \sqrt {\frac{Z}{{AN}}} \widehat {N{e_i}}} \right)} } \right]\left| {\Psi \left( 0 \right)} \right\rangle,
\end{array}
\end{equation}
with $\varepsilon$ being calculated by
\begin{equation}
\varepsilon  = \frac{{\Delta P}}{\hbar }\sqrt {\frac{A}{{NZ}}}.
\end{equation}
From the linear response theory, the response of dipole operator can be written as,
\begin{equation}
S\left( {\widehat R} \right) = \left\langle {\widehat R} \right\rangle \frac{\hbar }{{\Delta P\sqrt {A/\left( {NZ} \right)} }},
\end{equation}
\begin{equation}
- {\rm{Im}}S\left( {\hat R} \right) = \sum\limits_n {{{\left| {\left\langle n \right.\left| {\hat R} \right|\left. 0 \right\rangle } \right|}^2}\delta \left( {\omega  - {\omega _n}} \right)},
\end{equation}
where $n$ indexes different excited states.
Since evolution of the excited wave function to the final state can be obtained by the EQMD model,
Eq. \ref{eq:sum} can the sum rule,
\begin{equation}
\begin{array}{l}
\mathop \sum \limits_n \left[ {\left\langle n \right.|\hat R|{{\left. 0 \right\rangle }^2}\delta \left( {\omega  - {\omega _n}} \right)} \right] \\
=  - Im\frac{\hbar }{{\pi \Delta P\sqrt {A/(NZ)} }}\mathop \smallint \limits_0^\infty  \left\langle {\psi \left( t \right)} \right.|\hat R|\left. {\psi \left( t \right)} \right\rangle {e^{i\omega t}}dt.
\end{array}
\label{eq:sum}
\end{equation}
For E1 excitation, the cross section can be expressed as:
\begin{equation}
\sigma \left( \omega  \right) = 4\pi \alpha \frac{{\hbar \omega }}{{\Delta P\sqrt {A/(NZ)} }}\left( { - Im\mathop \smallint \limits_0^\infty  \left\langle {\psi \left( t \right)} \right.|\hat R|\left. {\psi \left( t \right)} \right\rangle {e^{i\omega t}}dt} \right),
\end{equation}
where $\omega$ is excitation energy, and $\alpha$ is the fine structure constant.

It is confirmed that the two ways for calculating GDR spectra come to the same result.
The following calculations are obtained by the response function method.
In EQMD calculations for dipole oscillations of light nuclei, the systems response is not of perfect linearity. The position of peaks of GDR spectra is dependent on oscillation energy (Fig.\ref{fig:peak}). The higher the excitation energy is, the lower the peak of GDR moves to. Since the excitation energy of GDR is usually in the range of 10 to 40 MeV,  the width of GDR spectrum shifting is about 2 MeV.  Because of nonlinearity of response which should be considered, we introduce a new normalization method (Eq. \ref{eq:norm}) to take the width into account.
\begin{equation}
{\frac{dP}{dE}}_{norm} = \frac{dP/dE}{\int_0^\infty (dP/dE) dE}.
\label{eq:norm}
\end{equation}
In realistic calculations, the normalized dP/dE is calculated in the excitation energy region of 8-35 MeV, which includes almost all the GDR peaks physically relevant.

In our calculations, no boundary of grids is used.
So the Fourier transform Eq.(\ref{eq:Fourier}) does not induce spurious effects \cite{Reinhard2006PRE73}. Because of the absence of decay channels, the damping of collective motions is not reasonable. In this context, the integration time of Fourier transform should be cut according to experiments and here 600 fm/c is a reasonable selection. The finite integration time will bring additional spreading to GDR spectrum, however, the spreading is less than 1 MeV, which is less than the width of GDR spectrum.

To get the accurate cross section of giant modes, high order effect beyond mean-field pairing correlations and a more accurate description of continuum states is needed. However, in our result, the dP/dE is arbitrary unit. A smoothing parameter $\Gamma = $ 2 MeV is applied, when the dP/dE spectrum is displayed.

\section{\label{sec:3}Results and discussion}
\subsection{\label{sec:3.1}$\alpha$ cluster in ground states}

\begin{figure}[H]
\centering
 \begin{overpic}[width=.8\textwidth]{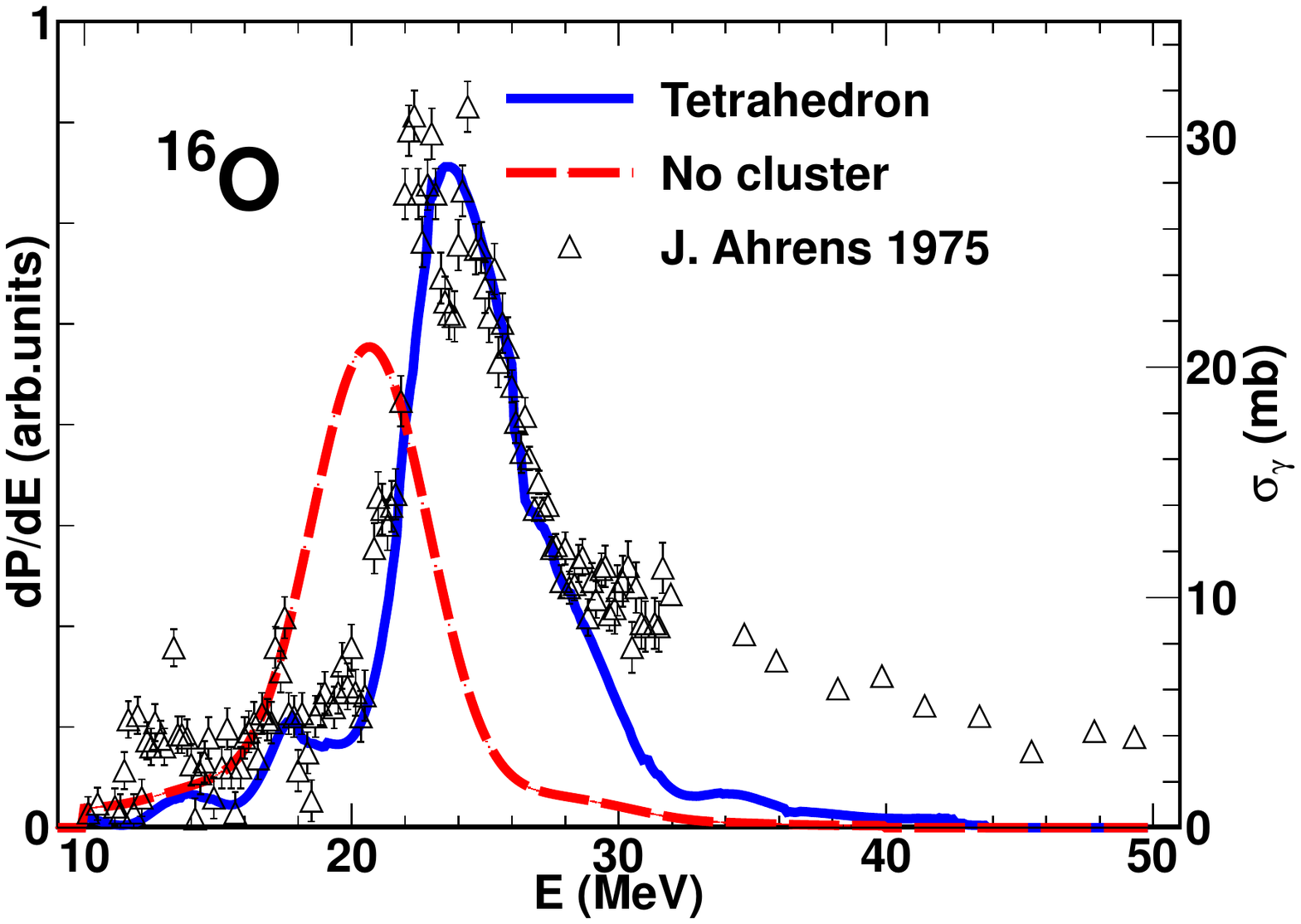}
 \end{overpic}
\vspace{-0.3cm}
\caption{\footnotesize (Color online) Comparison of the GDR calculation for $^{16}$O against experimental data (nuclear photoabsorption cross section on
the oxygen target) in Ref.\cite{Ahrens1975NPA251} (J. Ahrens 1975, empty triangles, scaled by the right Y axis). Solid blue line, tetrahedral $\alpha$-cluster state. Long dashed red line, non-cluster state.}
\label{fig:ground_o16}
\end{figure}

\begin{figure}[H]
\centering
 \begin{overpic}[width=.8\textwidth]{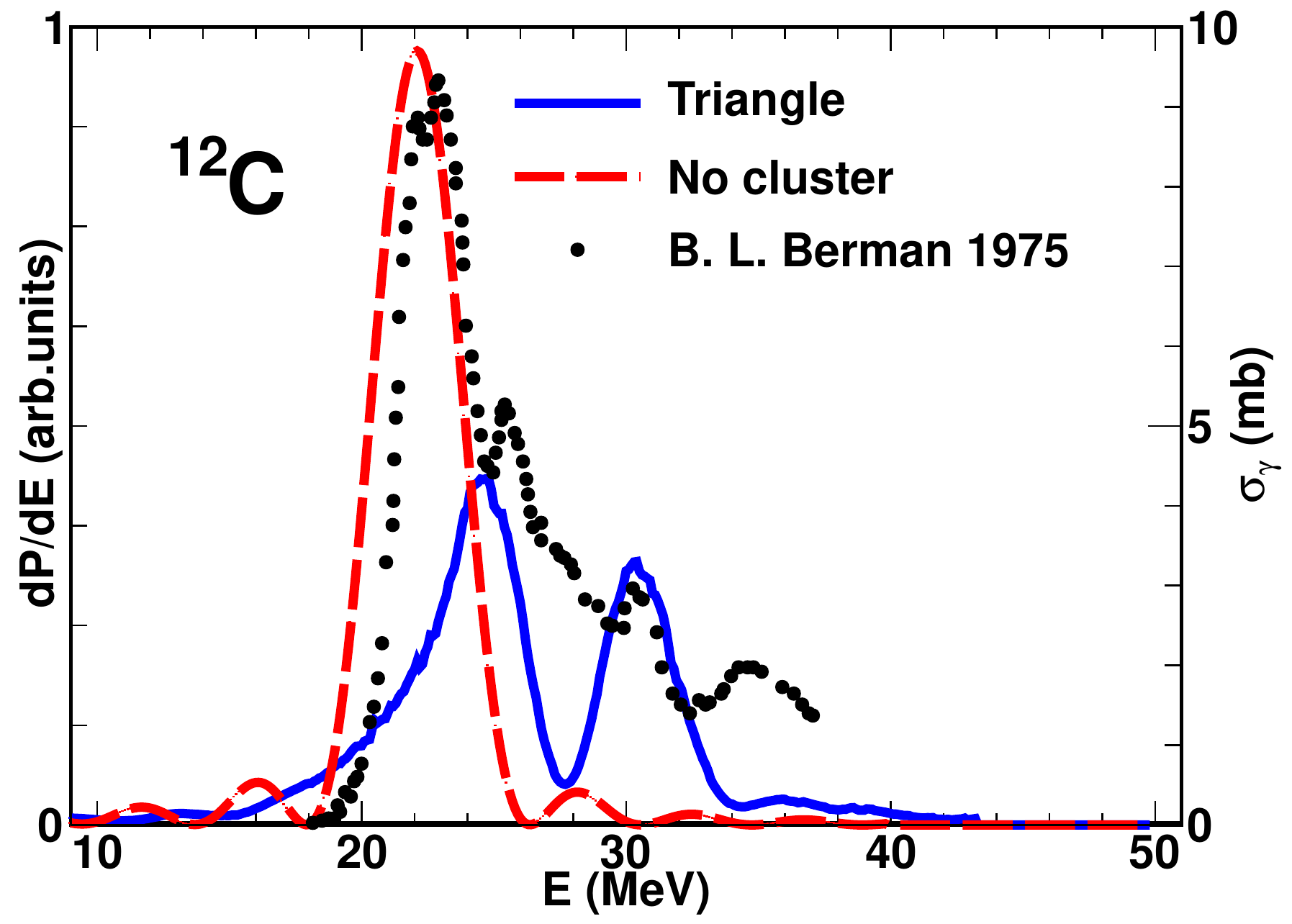}
 \end{overpic}
\vspace{-0.3cm}
\caption{\footnotesize (Color online) Comparison of the GDR calculation for $^{12}$C against experimental data  in Ref.\cite{Berman1975RMP} (B. L. Berman 1975, black dots, scaled by the right Y axis).
Solid blue line, triangle $\alpha$-cluster state.
Long dashed red line, non-cluster state.}
\label{fig:ground_c12}
\end{figure}

In EQMD model, $^{16}$O ground state is obtained at binding energy of 7.82$A$ MeV, which is close to the experimental binding energy: 7.98$A$ MeV, and consists of 4-$\alpha$ with a tetrahedral configuration. The tetrahedral 4-$\alpha$ configuration in $^{16}$O ground state is also supported by an {\it ab initio} calculation by Epelbaum, $et~al.$ \cite{Epelbaum2014PRL112} using nuclear chiral effective field theory.
A recent covariant density functional theory calculation also shows regular tetrahedral 4-$\alpha$ configuration in $^{16}$O ground state\cite{Liu2012CPC36}.
The non-cluster $^{16}$O ground state in EQMD can be obtained with the wave packet width of 4.2-4.3 fm for all the nucleons.
This width is much wider than that of nucleons in cluster states, in which all the nucleons have the width of just 1.9-2.1 fm. So the independent and variable wave packet width for each nucleon plays a crucial role on clustering, which is a distinct advantage of EQMD.
Fig.{\ref{fig:ground_o16}} shows the GDR results of non-cluster and cluster $^{16}$O ground states, together with the experimental data in Ref. \cite{Ahrens1975NPA251}.
The GDR of non-cluster ground state can not reproduce the data and the centroid is 4 MeV lower than the centroid of main peak of data. On the contrary, the GDR of tetrahedral configuration can reproduce the data well. So the tetrahedral 4-$\alpha$ configuration in initialization is reasonable and the procedure used to calculate GDR is reliable. For $^{12}$C, the non-cluster ground state is also obtained, in which the wave packet width of all the nucleons range from 3.5 to 3.6 fm. Fig.\ref{fig:ground_c12} shows a comparison between calculated result of $^{12}$C and data. The non-cluster ground state can reproduce the shape of low energy peak quite well with only about 1 MeV centroid shift. The centroid of high energy small peaks can be obtained from triangle $^{12}$C ground state. It is reasonable to infer the ground of $^{12}$C is a multi-configuration mixing of shell-model-like and $\alpha$ cluster configurations, which is consistent with the calculations of AMD \cite{Kanada-En'yo1998PRL81} and FMD \cite{Chernykh2007PRL98} models.


\subsection{\label{sec:3.2}$\alpha$ cluster configurations around threshold of n $\alpha$ breakup}

\begin{figure}[H]
\centering
\includegraphics[width=.9\textwidth]{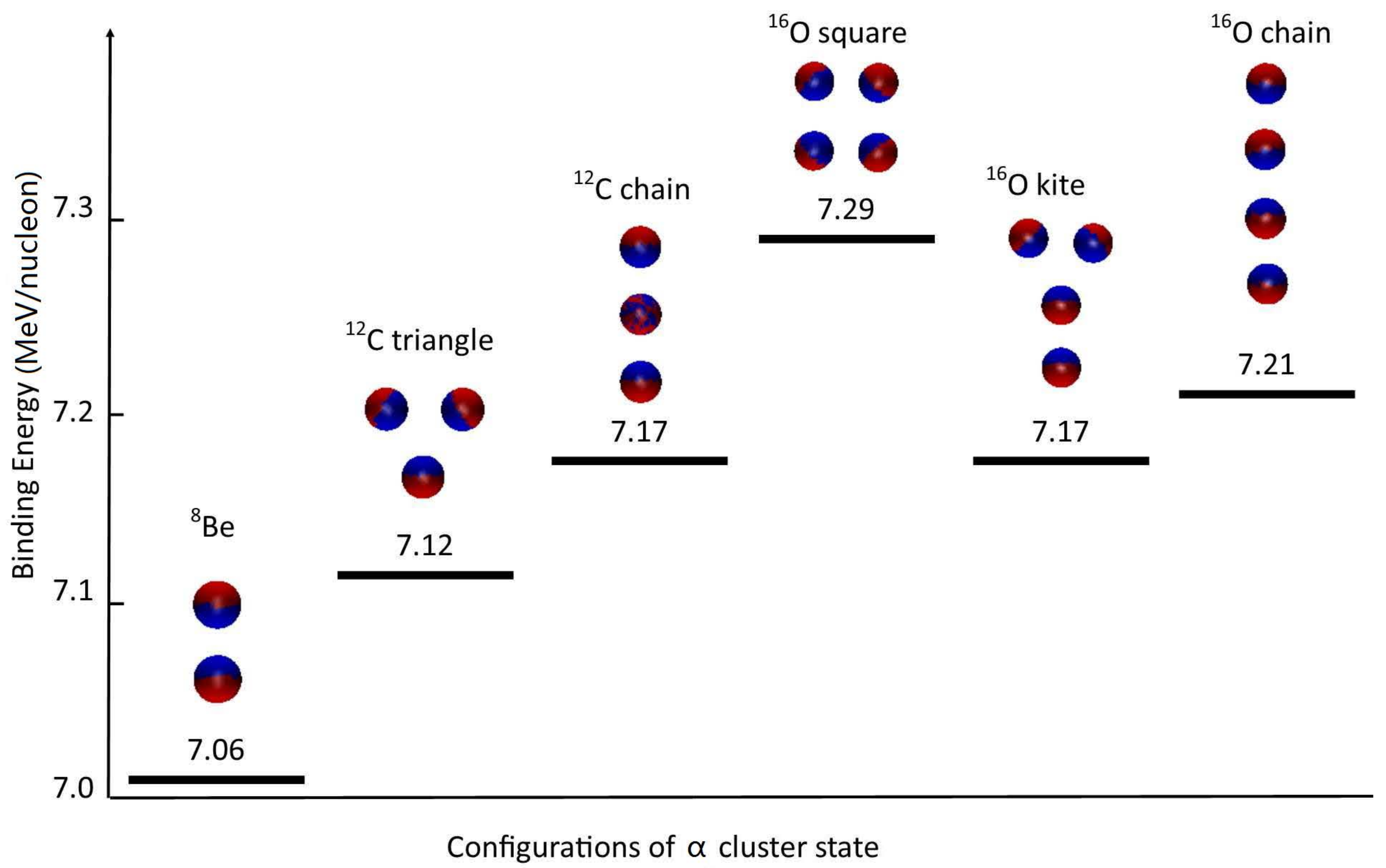}
\vspace{-0.3cm}
\caption{\footnotesize (Color online) Average binding energy for different $\alpha$ cluster structure around the threshold of n$\alpha$ breakup.}
\label{fig:alpha_binde}
\end{figure}

In EQMD model framework, $^8$Be at ground state has $\alpha$ cluster structure. It consists of two-alpha clusters, with 7.06 $A$ MeV binding energy. $^{12}$C has two possible $\alpha$ cluster structures (Fig.\ref{fig:alpha_binde}). One is triangle structure, with 3-$\alpha$ clusters forming a regular triangle shape. Its binding energy is 7.12$A$MeV, a little bigger than that of $^8$Be. The other is chain structure,in binding energy of 7.17$A$ MeV, which means that chain structure formed by three-alpha clusters is more stable than triangle structure. Similar to other theoretical predictions, $\alpha$ cluster states of light nuclei are shown up around the threshold energy to decay into free $\alpha$ particles. In EQMD result, $^8$Be is the closest to the threshold, and  other $\alpha$ cluster states inside heavier nuclei have bigger binding energy than the threshold. For a nucleus with different $\alpha$ cluster states, the binding energies of different cluster states differ very little from each other, which indicates different energy levels. The $^{16}$O $\alpha$ cluster states have three structures, with bigger binding energies (than those of $^8$ Be and $^{12}$C $\alpha$ cluster states), being 7.29, 7.17, and 7.21$A$ MeV for the square, kite, and chain structures respectively. Consequently, the most stable $\alpha$ cluster structure for $^{16}$O is square, and then the chain structure. Kite is the most unstable structure.
The excitation energy shown in Fig.\ref{fig:alpha_binde} for $^{12}$C and $^{16}$O is very near the predicted threshold energy in Ikeda diagram\cite{Ikeda1968PTPSE68}. And we have checked our results that they are not sensitive to the tiny binding energy difference, just sensitive to the geometric configurations of clusters.
\begin{figure*}[htbp]
\centering
 \begin{overpic}[width=.9\textwidth]{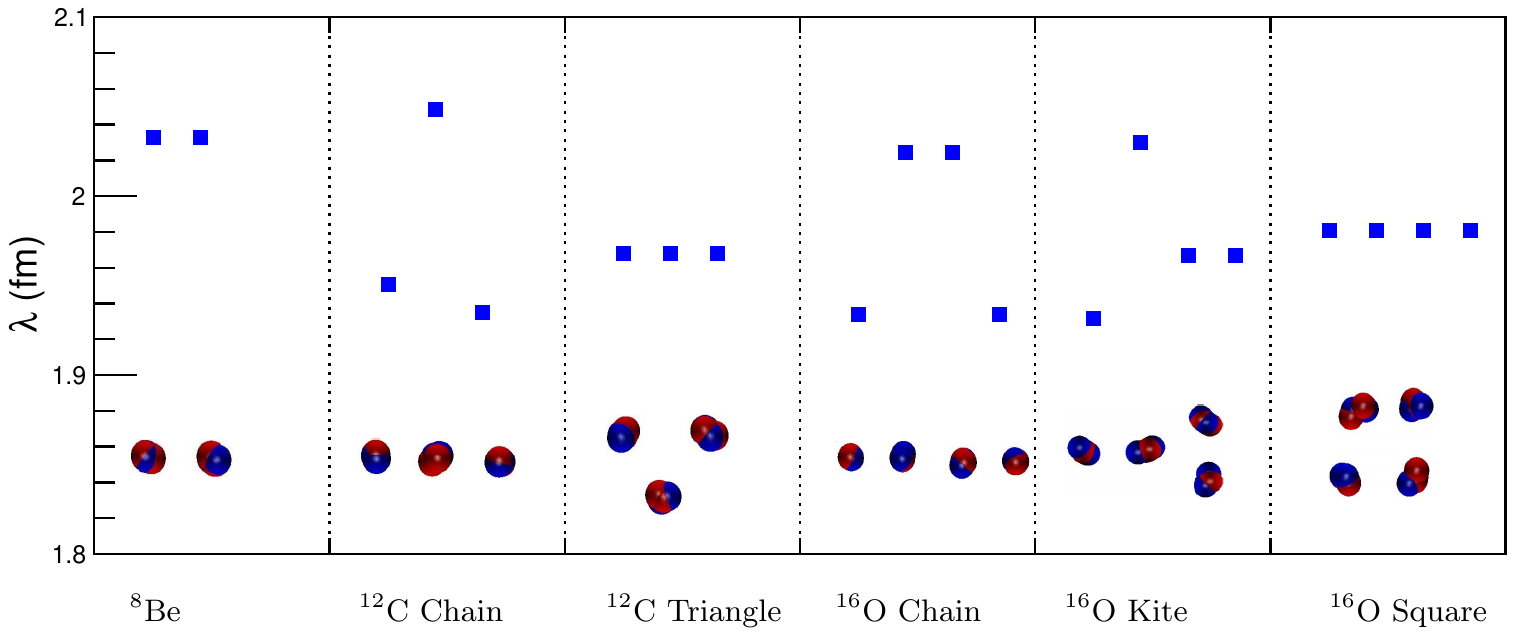}
  \end{overpic}
\caption{\footnotesize (Color online) Width of Gauss wave packets in different clustering states. Blue and red bubbles above horizontal axis are different configurations of $^8$Be, $^{12}$C, and $^{16}$O. Blue square marks are $\lambda$ of wave packets for different clusters, scaled by vertical axis.}
\label{fig:wlam}
\end{figure*}

Alpha cluster structures that are less symmetrical like $^{12}$C chain structure comparing to triangle structure, are more stable. This property indicates that $\alpha$ clusters in $^{12}$C chain structure state play different roles. For instance, the $\alpha$ cluster at the centre of $^{12}$C chain structure has larger Gauss wave packet width, which can help to hold the $\alpha$ clusters at both ends of the chain. The $\lambda_i$ of nucleons in different $\alpha$ clusters are shown in Fig.\ref{fig:wlam}.
As one can see in $^{16}$O chain structure and kite structure, $\alpha$ clusters at the centre of a nucleus  have larger Gauss wave packets width than outer ones.

\begin{figure}[H]
\centering
\includegraphics[width=.9\textwidth]{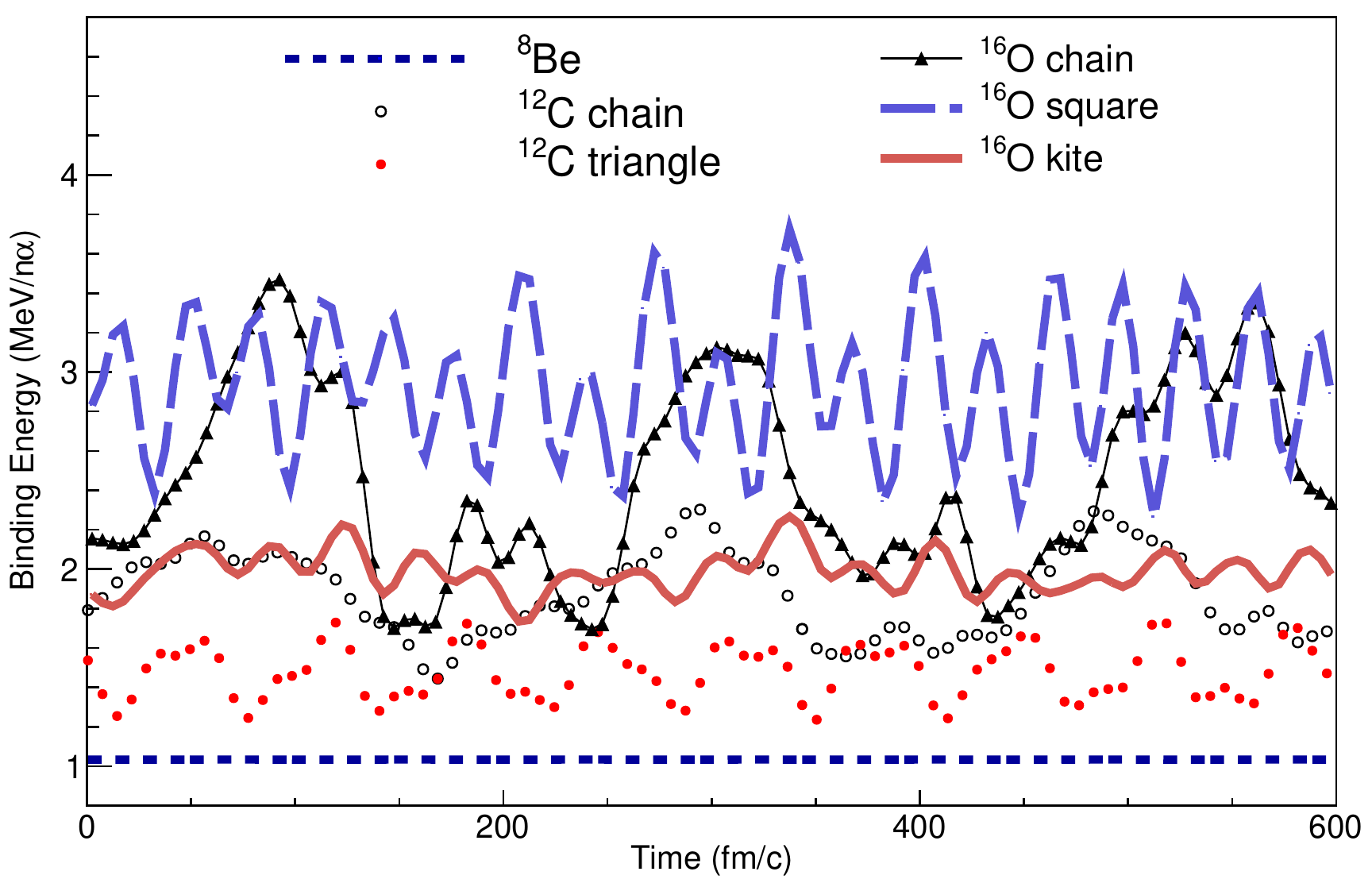}
\vspace{-0.3cm}
\caption{\footnotesize (Color online) Time evolutions of binding energy among $\alpha$ clusters for different configurations.}
\label{fig:evo}
\end{figure}

In EQMD framework, the binding energy of $\alpha$ clusters in nuclei is smaller than free $\alpha$ particles. For $^8$Be, the calculated internal binding energy of $\alpha$ cluster is 27.2 MeV, which is a little less than the experimental result (28.3 MeV) of free $\alpha$ particle. Considering the binding energy of $^8$Be system is 7.06$A$ MeV, one knows that the binding energy between two alpha-clusters is 1.02 MeV/$\alpha$. For $^{12}$C and $^{16}$O, the binding energy among $\alpha$ clusters show larger value. Under the time evolution, those binding energy will oscillate periodically (see Fig.\ref{fig:evo}). In fact, $^8$Be binding energy between two $\alpha$ clusters oscillates, in very little amplitudes ($<$ 0.01 MeV), though. Fig.\ref{fig:evo} shows that the periods are very different and sensitive to $\alpha$-cluster structure, and every oscillation consists of multiple frequencies. For $^{16}$O chain structure, more than two periods are with difference of $>$ 200 fm/c. The oscillation of binding energy indicates that energy flows into and go out the $\alpha$ clusters periodically.

\begin{figure}[H]
\centering
\includegraphics[width=.9\textwidth]{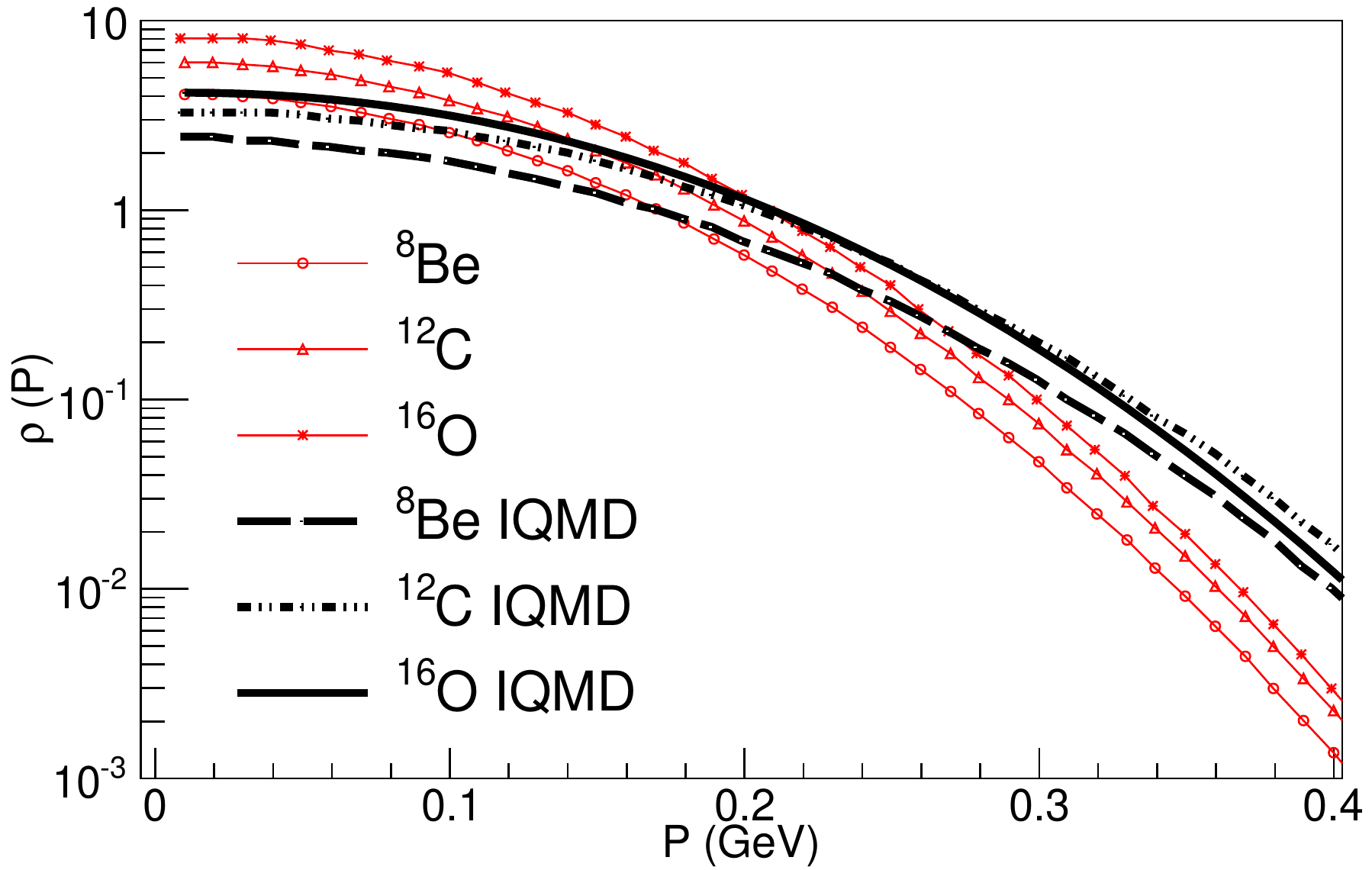}
\vspace{-0.3cm}
\caption{\footnotesize (Color online) Momentum distributions of different nuclei, in EQMD model and standard isospin-dependent quantum molecular dynamics(IQMD) model. Black lines show momentum distributions in a standard IQMD model, and red line with marks represent EQMD model's results.}
\label{fig:momentum}
\end{figure}

The momentum distribution of cluster states differ greatly from normal nuclei at ground state. Fig.\ref{fig:momentum} shows momentum distributions of $^8$Be, $^{12}$C, and $^{16}$O. The red data point are clustering state results calculated in EQMD model, where the black lines are the non-clustering state results calculated in IQMD model. The momentum distribution is not sensitive to different structures of $\alpha$ clusters. As shown by the red marked lines, $^{12}$C chain structure and triangle structure give the same result, and $^{16}$O chain, square, and kite structures also give the same results. One can see from Fig.\ref{fig:momentum} that at low momentum region, clustering nuclei have higher value of momentum distributions than non-clustering nuclei, while this reverses at high momentum region.

To calculate GDR spectrum, one can give the nucleus a boost to obtain the dipole oscillation, or simulate a Coulomb excitation with a heavier nucleus. The two methods give the same result. But the first method gives no information about the difference between $\alpha$ clusters in a nucleus. To discuss the dipole motion's coherence of different $\alpha$ clusters in a nucleus, the following results of this section are obtained by simulations of clustering-nuclei as projectiles to hit $^{40}$Ca as target. In details, the impact parameter is 20 fm, and the projectiles are 100 MeV in incident energy. The systems evolve stop at 600 fm/c. The length of time will affect the GDR spectrum width in EQMD calculations. The shorter the calculation time, the wider the spectrum are obtained. Then, the time should be grater than 300 fm/c, so as not to come up with a too wide GDR spectrum width. Because the oscillations excited by Coulomb reaction are of small amplitude, the peaks of GDR spectra in this article shift 1 MeV towards high energy compared with our previous results \cite{He2014PRL113}.
Another point should be mentioned here is that the excited energy of the following results are based on cluster states which are all low-lying states and different from the excited energy mentioned in the section \ref{sec:3.1}(Fig.\ref{fig:ground_o16} and Fig.\ref{fig:ground_c12}) which are based on ground states.

\subsection{$^8$Be dipole oscillation}
\begin{figure*}[htbp]
\flushleft
 \begin{overpic}[width=\textwidth]{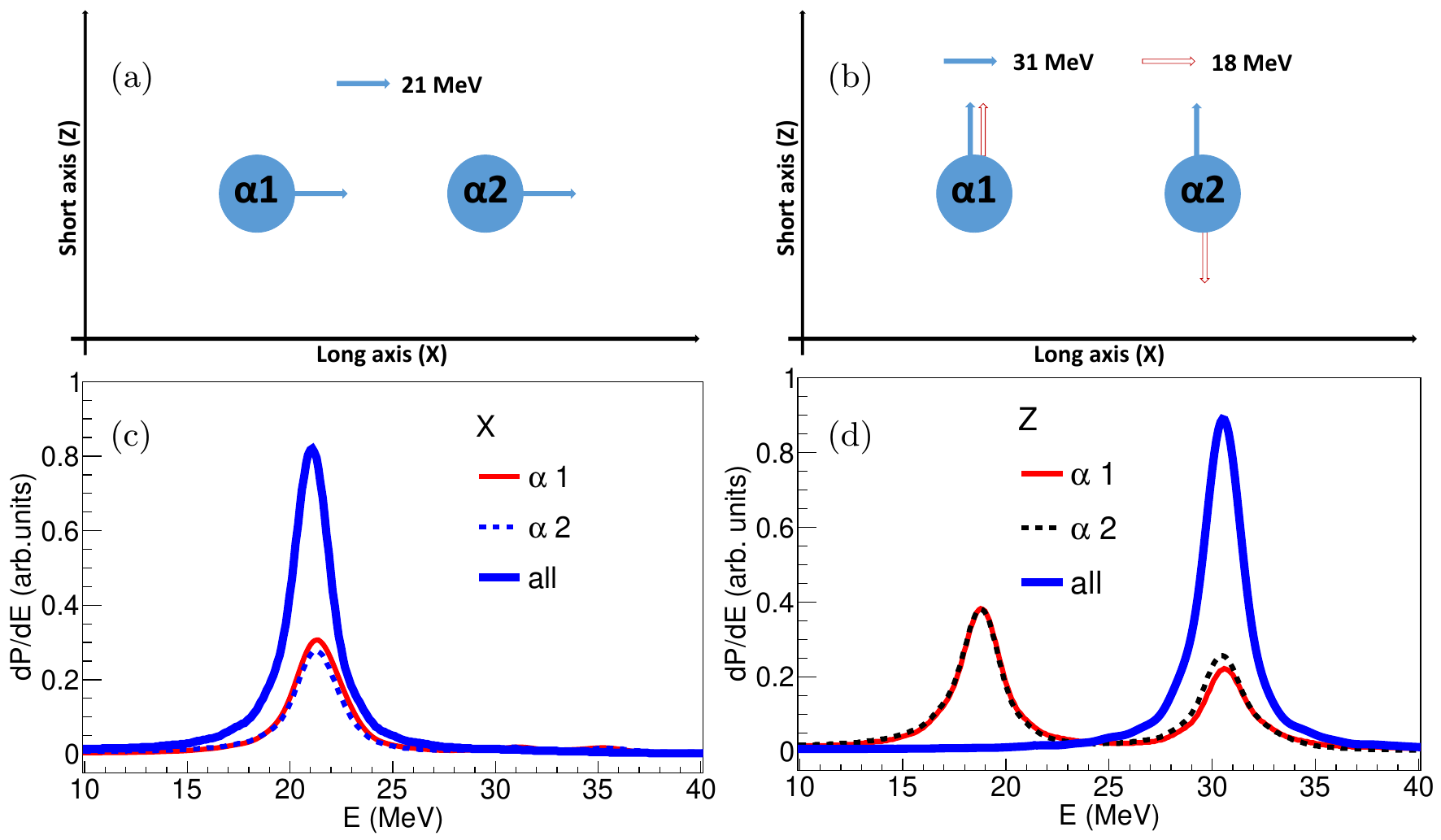}
  \end{overpic}
\caption{\footnotesize (Color online) Coherence of $^8$Be dipole motion. (a) and (c) show the coherence of motion along the long axis. (b) and (d) show the coherence of motion along the short axis. Arrows in (a) and (b) indicate the phase of oscillations in different frequencies.}
\label{fig:coherent_be8}
\end{figure*}

As a collective motion of nucleons, GDR will be affected sensitively by nuclear exotic structures. $\alpha$ clusters will split and complicate the GDR spectrum. To discuss how $\alpha$ clusters affect collective motions, $^8$Be is the simplest example. Fig.\ref{fig:coherent_be8} shows the GDR spectrum of $^8$Be. The two peaks at 21 MeV and 31 MeV are contributed by the long and the axis of system, respectively. There exists a rule that the dipole oscillation frequency in any direction of a cluster system is inversely-proportional to the length of system's configuration. In Fig. \ref{fig:coherent_be8}, both alpha-clusters have a single frequency in long axis direction at 21 MeV with the same oscillation phase.
It should be noted that the arrows drown on the clusters do not indicate the motion direction of the whole $\alpha$ cluster, but mean the iso-vector dipole motion in the $\alpha$ cluster, in which the two protons move against the two neutrons. For example, in Fig.\ref{fig:coherent_be8}(B), the two filled blue arrows with the same direction mean that the two $\alpha$ clusters have the same direction of iso-vector dipole motion with the same oscillation phase, the two empty red arrows with opposite direction mean that the two $\alpha$ clusters have the same direction of iso-vector dipole motion but with the opposite oscillation phase.
For this case, in short axis direction, every $\alpha$ cluster has two frequencies at 31 and 18 MeV. The oscillations of two alpha-clusters at 31 MeV are coherent with the same phase, while the peak at 18 MeV disappears when two alpha-clusters are considered as a whole system. The oscillations with this frequency are coherent with opposite phase.

\subsection{$^{12}$C dipole oscillation}
\subsubsection{$^{12}$C triangle structure}

\begin{figure}[H]
\centering
  \begin{overpic}[width=0.6\textwidth]{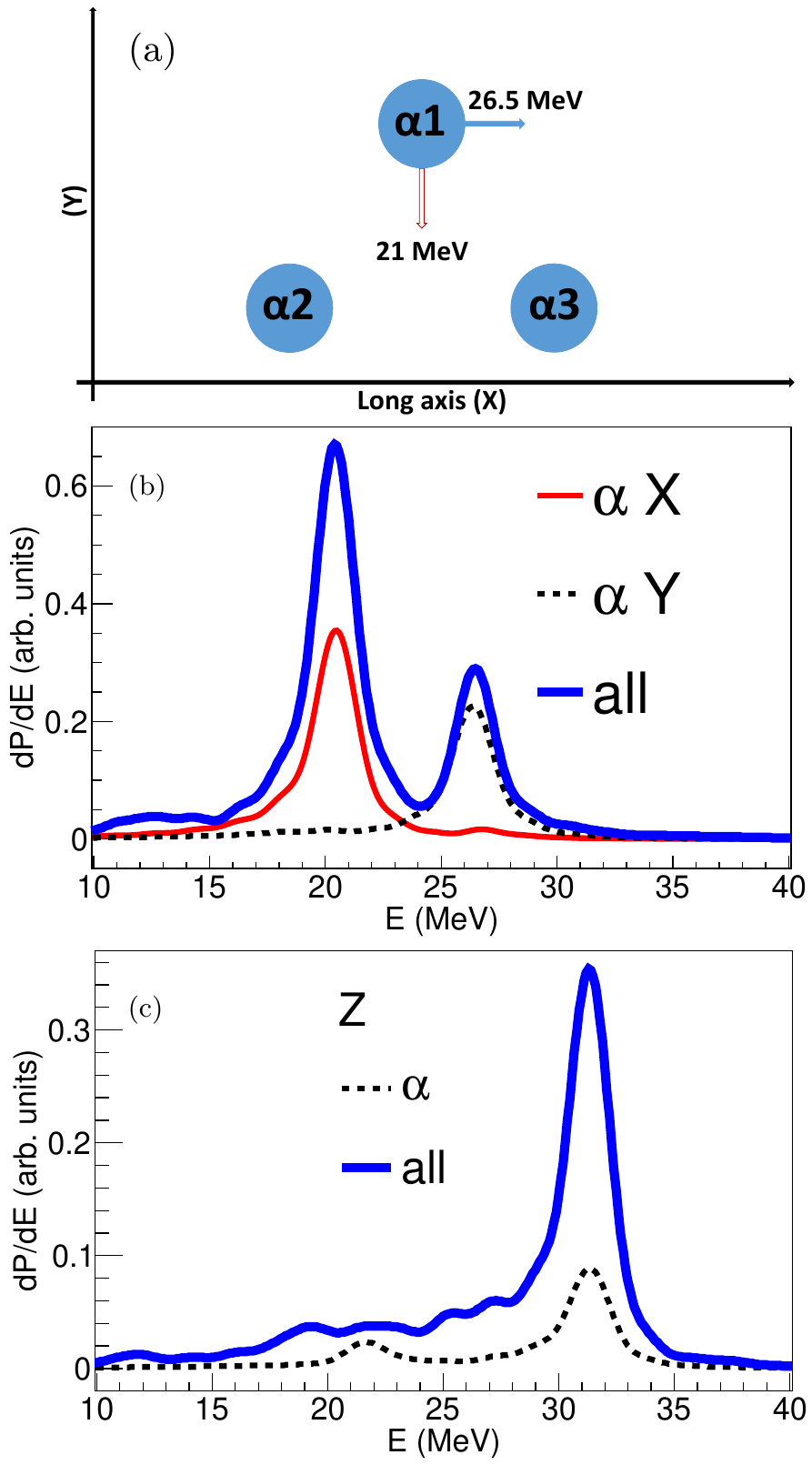}
  \end{overpic}
\caption{\footnotesize (Color online) Coherence of $^{12}$C triangle structure dipole motion. (a) and (b) show the coherence of motion along the long axis. (c) shows the coherence of motion along the short axis. Arrows in (a) indicate the phase of oscillations in different frequencies.}
\label{fig:coherent_c12_triangle}
\end{figure}

$^{12}$C GDR spectrum with triangle structure gives three peaks at 21, 26.5, and 31 MeV. For this configuration, the short axis is perpendicular to the plane determined by the triangle shape. In this direction, every $\alpha$ cluster has a main peak at 31 MeV, and a small peak at 21.5 MeV. But the little peaks of different $\alpha$ clusters are non-coherent. So this peak does not show up in the whole system GDR spectrum. The frequency of 31 MeV is coherent, which gives a strong peak. In long axis, every $\alpha$ cluster has two peaks at 21 and 26.5 MeV. The 21 MeV peak is close to GDR peak of $^8$Be in long axis and is supposed to be built by interaction of two alpha-clusters. To check the components of peaks at 26.5 MeV, the long axis is rotated by $\pi/6$. As shown in Fig.\ref{fig:coherent_c12_triangle}, after rotation, the two components of 26.5 and 21 MeV are separated. The oscillating direction of 26.5 MeV frequency is parallel to the bottom line of triangle shape. This frequency is proposed to be built by interaction among three alpha-clusters.

\subsubsection{$^{12}$C chain structure}
\begin{figure*}[htbp]
\flushleft
  \begin{overpic}[width=.9\textwidth]{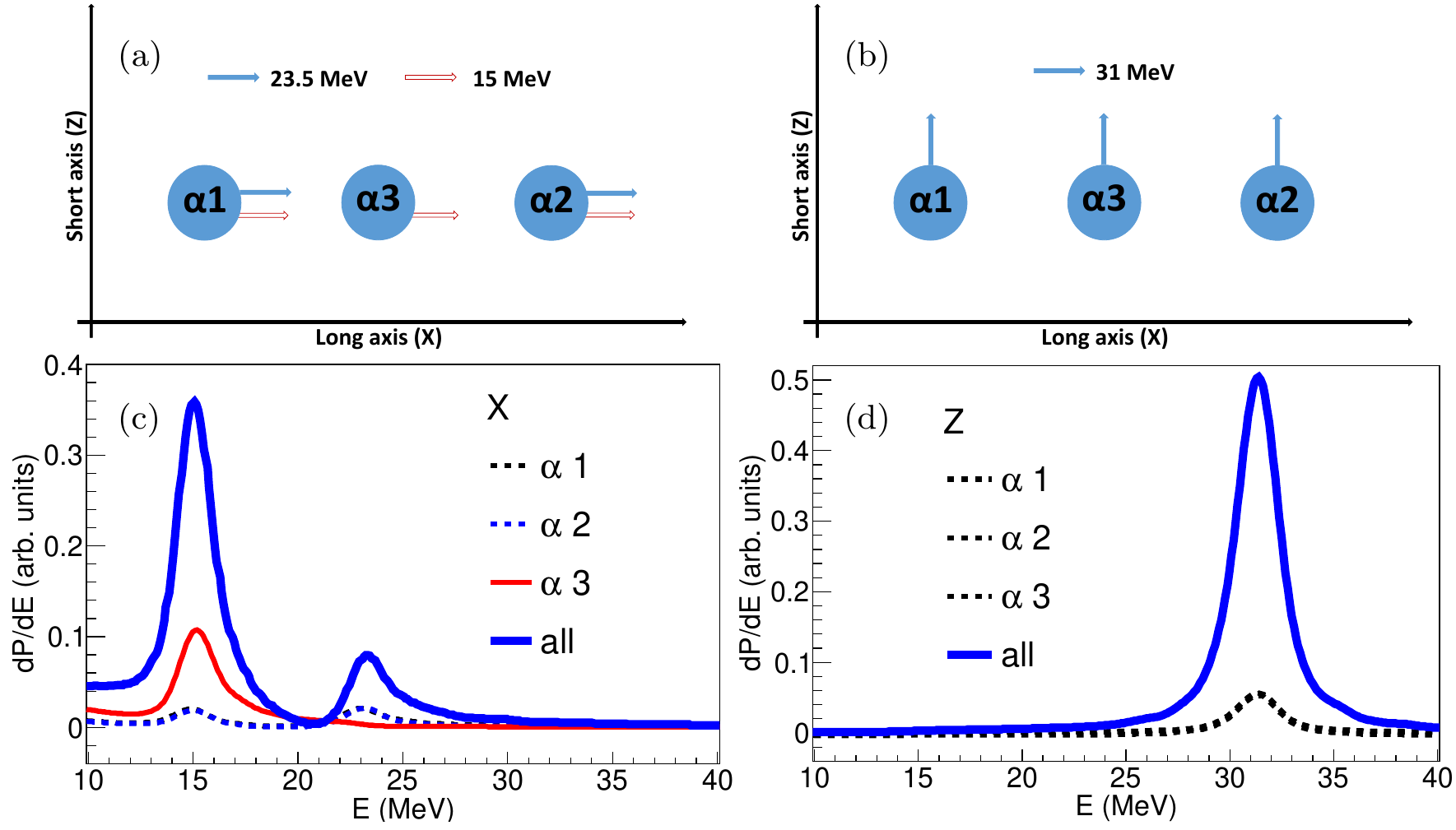}
   \end{overpic}
\caption{\footnotesize (Color online) Coherence of $^{12}$C chain structure dipole motion. (a) and (c) show the coherence of motion along the long axis. (b) and (d) show the coherence of motion along the short axis. Arrows in (a) and (b) indicate the phase of oscillations in different frequencies.}
\label{fig:coherent_c12_chain}
\end{figure*}

The results of $^{12}$C with chain structure are shown in Fig.\ref{fig:coherent_c12_chain}, with three components at 15, 23.5, and 31.5 MeV. The 31.5 MeV peak is along the short axis, and is contributed equally by every $\alpha$ cluster, while the 15 MeV mainly by the central $\alpha$ cluster. The $\alpha$ clusters at both ends contribute to it, weakly though. The central $\alpha$ cluster feels stronger 3-$\alpha$-interaction than two other-side $\alpha$ clusters. For the peak at 23.5 MeV, it is supposed to be built by 3-$\alpha$-interaction like triangle structure. This kind of 3-$\alpha$-interaction is sensitive to the configuration of $\alpha$ clusters. The regular triangle structure shows stronger oscillation strength at higher frequency of 26.5 MeV, and chain structure gives weak strength and lower frequency at 23.5 MeV. Because of the weak oscillations caused by Coulomb excitation, it does not show up peak around 21 MeV, which indexes the two-alpha interaction. In fact, when the dipole oscillation has larger strength, the triangle structure may degenerate into a $^8$Be substructure and an $\alpha$ bounded to it very weakly. Then it will give a peak around 21 MeV.

\subsubsection{$^{16}$O chain structure}
\begin{figure*}[htbp]
\flushleft
    \begin{overpic}[width=.9\textwidth]{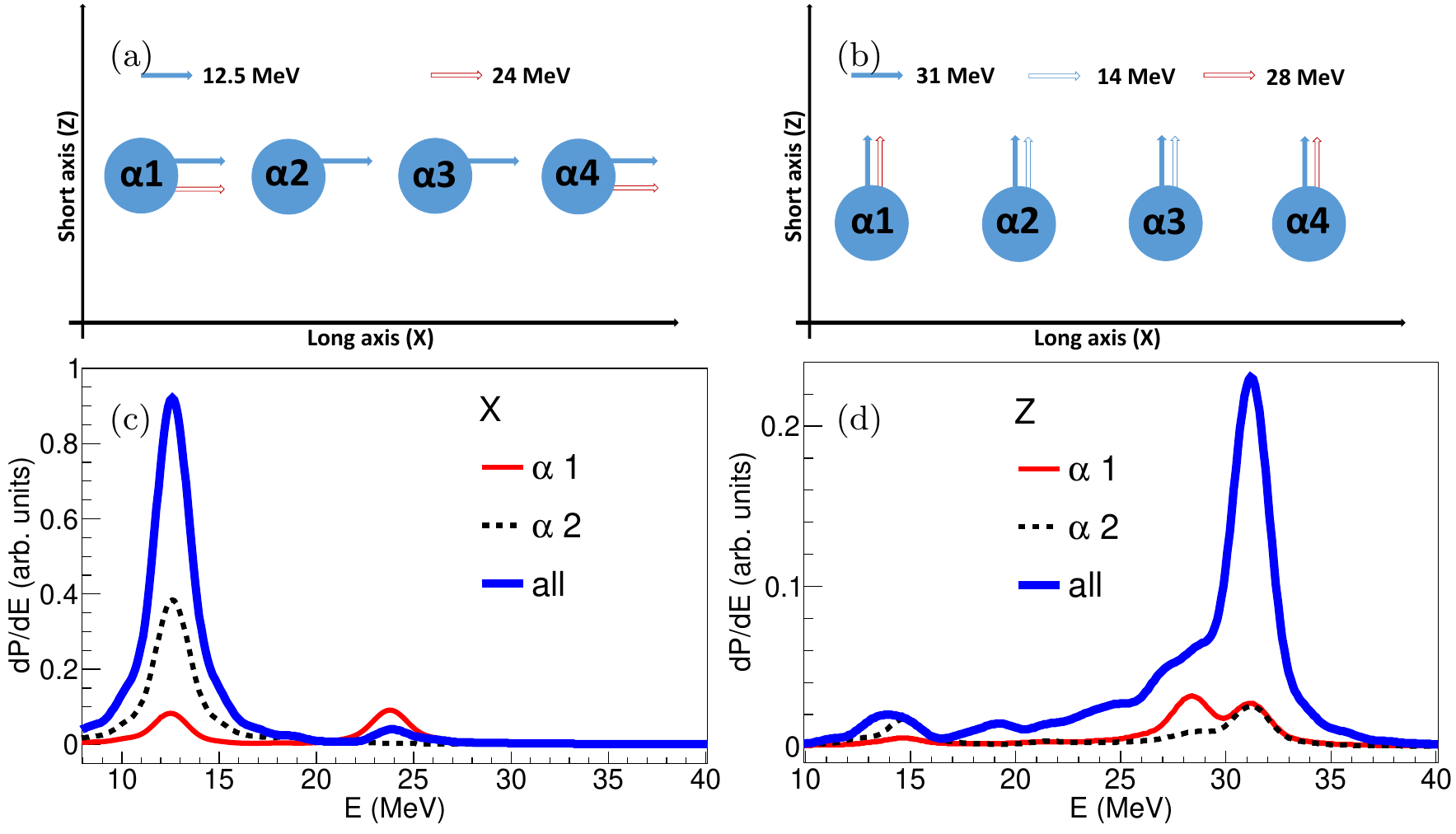}
    \end{overpic}
\caption{\footnotesize (Color online) Coherence of $^{16}$O chain structure dipole motion. (a) and (c) show the coherence of motion along the long axis. (b) and (d) show the coherence of motion along the short axis. Arrows in (a) and (b) indicate the phase of oscillations in different frequencies.}
\label{fig:coherent_o16_chain}
\end{figure*}

\begin{figure}[htbp]
\centering
\begin{overpic}[width=0.8\linewidth]{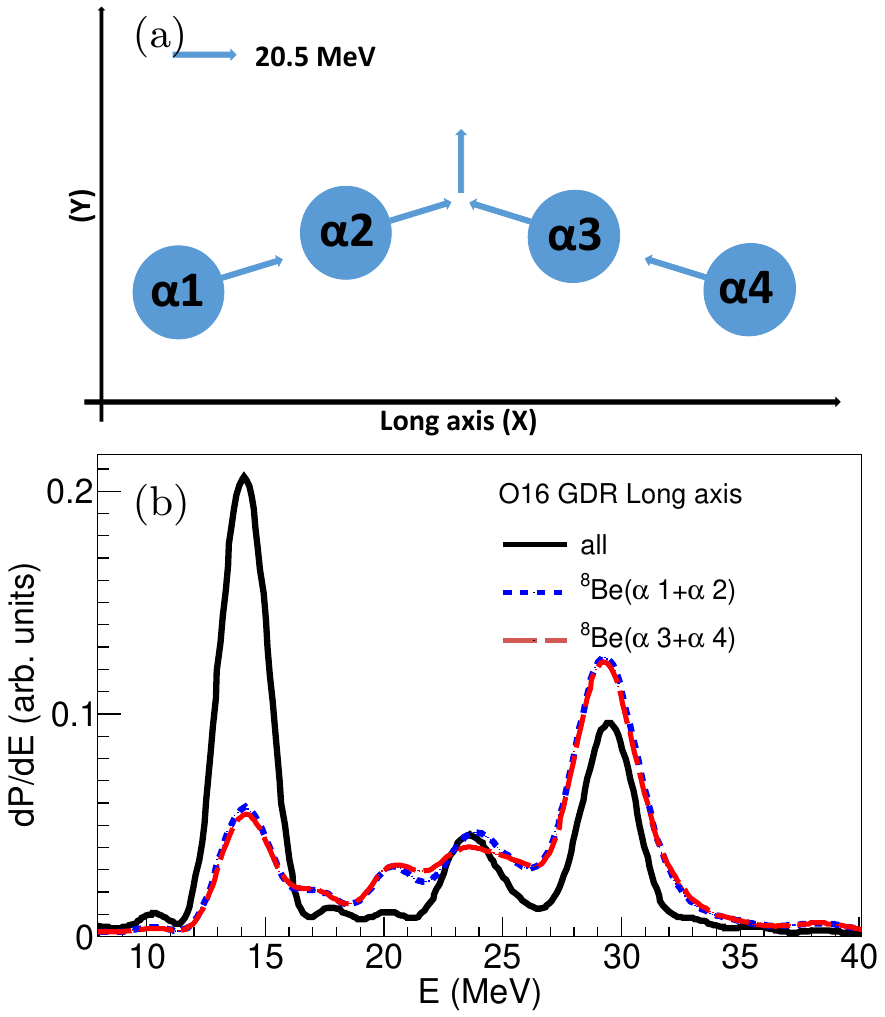}
\end{overpic}
\caption{(Color online) $^8$Be substructure in $^{16}$O chain state. (a) gives the phase of oscillation and (b) shows the GDR spectra along the long axis. The peak at 20.5 MeV raised by substructure of $^8$Be (long dashed line and short dashed line) are subtracted coherently, so the total spectra (solid line) do not have peak at 20.5 MeV.}
\label{fig:coherent_o16_chain_be8}
\end{figure}

Fig.\ref{fig:coherent_o16_chain} shows the spectrum of short axis. The coherent peak locates at 31 MeV, too. The $\alpha$ clusters at the ends of chain structure shows up a frequency at 28.5 MeV, which is supposed to be a multi-$\alpha$-interaction effect. As the previous results do not show up this peaks, it can be supposed to show up only when the oscillation strength is weak enough and the structure nearly keep regular linear chain. The same case happens at 14 MeV. It is a very small peak, and the oscillation are non-coherent, so its total strength is close to one $\alpha$ cluster. For the long axis, the main peak locates at 12.5 MeV, due to the mean filed effect of the system. Its size is the largest of all nuclei and structures, hence the lowest frequency of all. Like the case of $^{12}$C chain structure, the $\alpha$ clusters at the center have strongest oscillation strength. The peak at 24 MeV indicates that $\alpha$ clusters at the chain ends feel a three-alpha-interaction like $^{12}$C chain and triangle structures. But oscillation of the two $\alpha$ clusters is non-coherent, so the total GDR strength is not stronger than one $\alpha$ cluster. Fig.\ref{fig:coherent_o16_chain_be8} shows that the GDR spectrum of $^{16}$O chain structure is a little bent. If the chain is considered as two $^8$Be subsystems, each $^8$Be will show up a peak around 20.5 MeV in GDR spectrum. But the two subsystems are coherent with opposite phase, so the peak disappears in total GDR spectrum.

\subsubsection{$^{16}$O kite structure}
\begin{figure*}[htbp]
\flushleft
    \begin{overpic}[width=\textwidth]{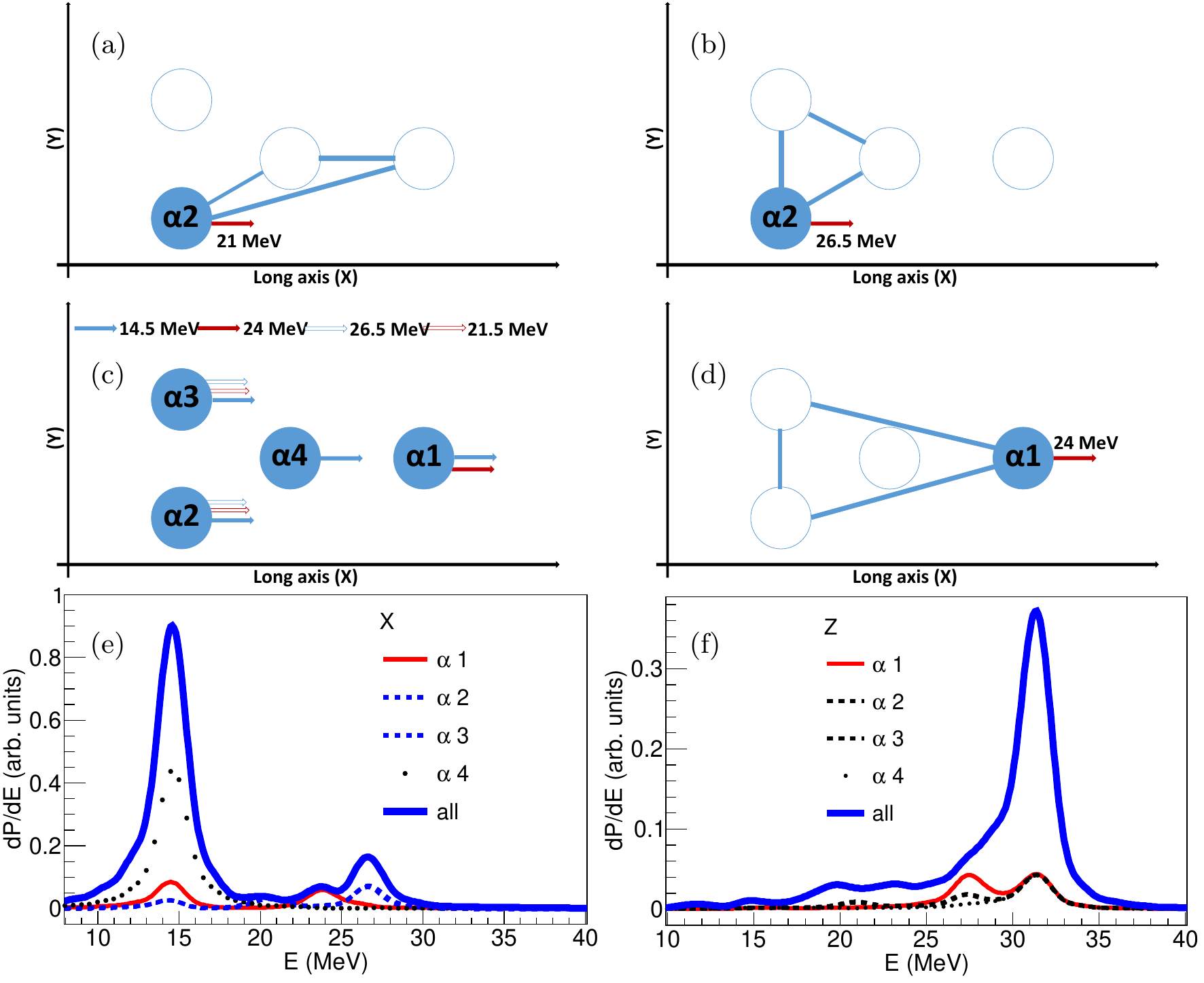}
    \end{overpic}
\caption{\footnotesize (Color online) Coherence dipole motion of $^{16}$O kite structure. (a), (b), $\&$ (d), different oscillations caused by three $\alpha$ substructure, with small peaks at 20-30 MeV. (c) $\&$ (e) coherence along the long axis. (f), GDR spectra along the short axis.}
\label{fig:coherent_o16_kite}
\end{figure*}

$^{16}$O kite structure is the most complicated $^{16}$O structure, and is more complicated than the $^{12}$C cluster states discussed above. It can be regarded as one triangle $^{12}$C added with a weakly bound $\alpha$ located at one of the vertex of triangle structure. The wave packets of nucleons in the central $\alpha$ cluster is larger than nucleons in other $\alpha$-clusters, including the $^{12}$C triangle state. This is the way to keep the additional $\alpha$ cluster bounded. Fig.\ref{fig:coherent_o16_kite} shows the GDR spectrum in short axis, all peaks are similar to the $^{16}$O chain structure, without the peak at 14 MeV. In the long axis, each $\alpha$ cluster behaves very differently. The GDR peak at 14.5 MeV is determined by the mean filed effect, with the central $\alpha$ cluster contributing the biggest of all the $\alpha$ clusters in triangle subsystem. In other words, the weakly bounded $\alpha$ cluster's contribute the lowest. The 3-$\alpha$-interaction is dependent on the structures of the related $\alpha$ clusters. The $^{16}$O kite structure can be decomposed into three kinds of triangle structures. As shown in Fig. \ref{fig:coherent_o16_kite}. $\alpha$ 1 gives a peak at 24 MeV, which is the effect of triangle 1 structure. This frequency is lower than that of the $^{12}$C triangle state, and higher than that of the $^{12}$C chain state, due to difference of the triangle structures. $\alpha$ 2 and 3 give peaks at 26.5 and 21.5 MeV, respectively. The 26.5 MeV peak is similar to the $^{12}$C triangle state, because of is the similar triangle structures. $\alpha$ 2(or 3), $\alpha$ 4, and $\alpha$ 1 make up an obtuse triangle structure. This special structure makes $\alpha$ 2 and 3 give a peak at lower energy, so the peak at 21.5 MeV is not caused by $^8$Be subsystem.

\subsubsection{$^{16}$O square structure}
\begin{figure}[htbp]
\centering
    \begin{overpic}[width=0.6\textwidth]{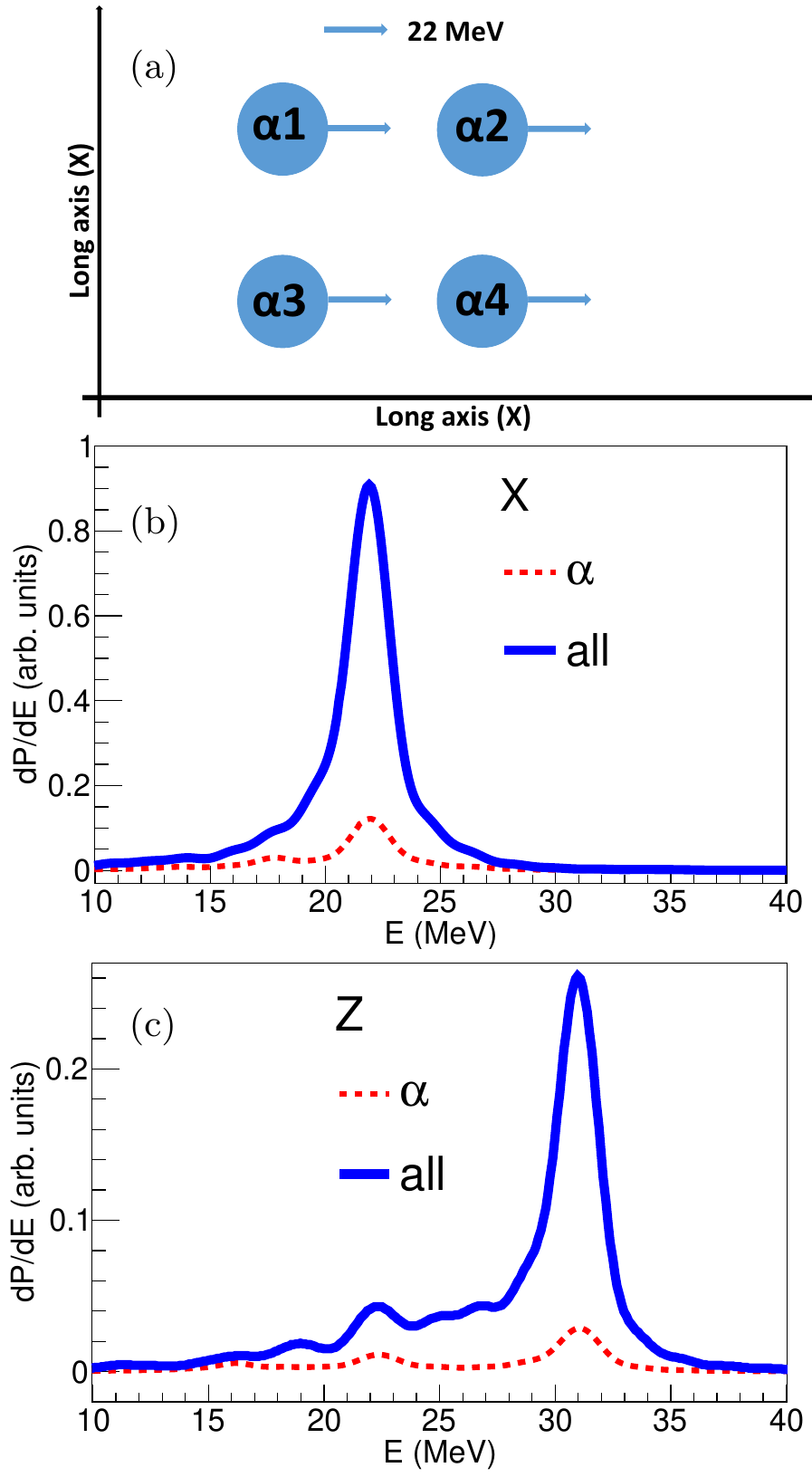}
    \end{overpic}
\caption{\footnotesize (Color online) Coherence of $^{16}$O square structure dipole motion. (a) $\&$ (b) the coherence of motion along the long axis. (c) coherence of motion along the short axis. Arrows in (A) indicate the phase of oscillation.}
\label{fig:coherent_o16_square}
\end{figure}

$^{16}$O square structure is a comparatively simple structure. As shown in Fig.\ref{fig:coherent_o16_square}, the short axis peaks is at 31 MeV, and the long axis peak, at 22 MeV. Because this state has the highest binding energy and the structure enhances three-alpha-interaction, the energy peak in long axis is the highest of all states.

\subsection{The difference of GDR spectrum between $\alpha$ cluster states and deformation}

\begin{figure}[htbp]
\flushleft
    \begin{overpic}[width=\textwidth]{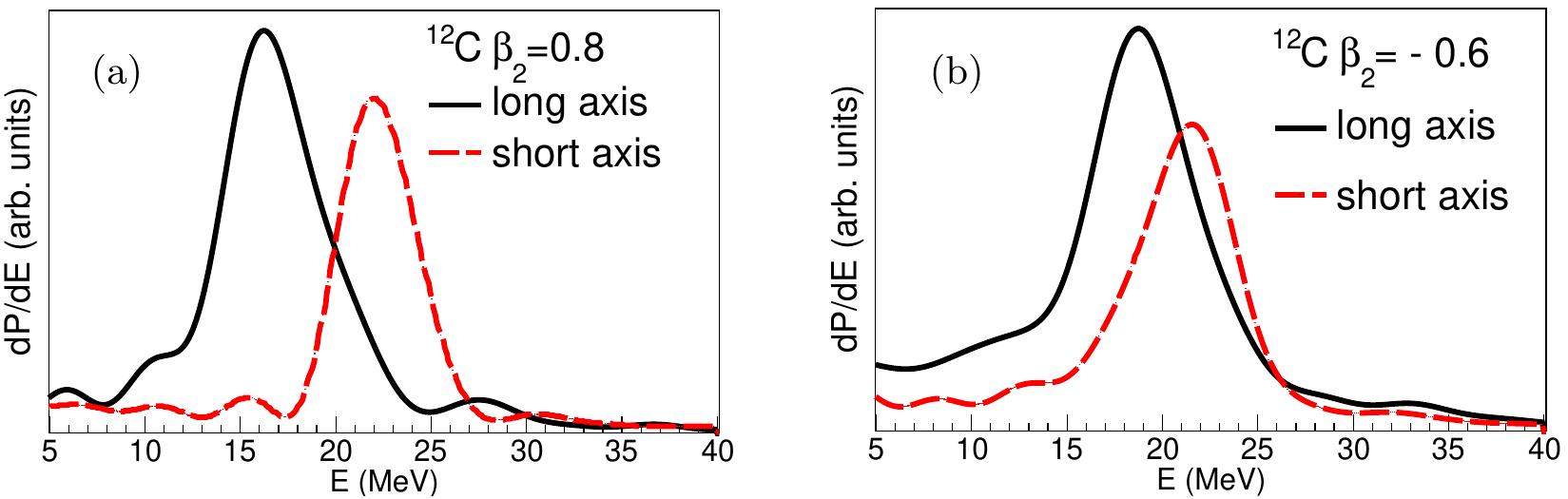}
    \end{overpic}
\caption{(Color online) GDR spectra split by deformation. (a) deformation with $\beta_2$=0.8; (b) deformation with $\beta_2$= -0.6. }
\label{fig:deformation}
\end{figure}

 The calculated GDRs of $^{12}$C in positive or negative $\beta_2$, without cluster, are shown in Fig.\ref{fig:deformation}. However, these $^{12}$C samples are initialized manually. The split between two peaks is much smaller than the split of GDR with cluster. And the missing of 30 MeV peak is a significant difference in comparison with the $\alpha$ cluster states. This is true for both prolate and oblate deformations.

The homogeneous prolate or oblate ellipsoid, which has two collective structure degrees of freedom, usually can have two obvious peaks, corresponding to the long and short axes, respectively, whereas a clustering GDR spectrum with multi-peaks has several collective degrees of freedom. The fragmented response function is a complicated coherent result of strong interference between alphas, which counteract or strengthen with each other. The GDR peak around 30 MeV is a reliable proof to confirm the existence of $\alpha$-clusters in light nuclei.

There are obvious substructure in GDR, such as similar GDRs of $^8Be$ and triangle $^{12}$C, appear as substructure in GDRs of chain $^{12}$C and kite $^{16}$O, respectively. This distinct feature of GDR can be taken as a strong signal which is different from the normal prolate or oblate deformation in light nuclei.

\section{conclusions}

In conclusion, the advantage and results of EQMD model to describe cluster states, and calculation method of isovector nuclear GDR, are discussed. Properties of ground states and $\alpha$ cluster states in light nuclei are discussed with an extended QMD model. $^8$Be at ground state consists of two $\alpha$ clusters. For $^{12}$C and $^{16}$O cluster states, clusters form different configurations. The average binding energy of $^{12}$C and $^{16}$O cluster states, around threshold energy for the decay into free $\alpha$ particles, are little bigger than $^8$Be at 7.07$A$ MeV. For cluster states, binding energy between $\alpha$ clusters are studied, which oscillates in several frequencies. So, energy flows into and goes out the clusters periodically. The dynamic Gaussian wave packets width is important for a QMD model to describe nuclear $\alpha$ cluster states. For different $\alpha$ cluster in a nucleus, the central ones have larger width, and the outer ones have smaller width. The momentum distribution of $\alpha$ cluster states differs greatly from the nuclei in standard QMD model.

Collective excitation of $\alpha$ cluster states shows interesting phenomena. The giant dipole resonances depend on geometric configuration of $\alpha$ clusters in a nucleus. Coulomb excitation process is applied to study the resonance rule of different $\alpha$ clusters. The dipole oscillation frequency in any direction of a clusters system is inversely-proportional to the length of system configuration. As the simplest system of cluster state, $^8$Be has one GDR peak at 31MeV oscillating perpendicular the long symmetrical axis of two $\alpha$ clusters and another peak at 21MeV oscillating along the long axis. The complicated behaviour of collective oscillations for $^{12}$C and $^{16}$O $\alpha$ cluster states with different configurations, are discussed in details. These show that GDR spectra of clustered nuclei are sensitive to $\alpha$ cluster states and configurations.

\begin{acknowledgments}

This work was supported in part  National Natural Science Foundation of China under contract Nos. 11421505, 11305239,
and 11220101005,  and the Major State Basic Research
Development Program in China under Contract No. 2014CB845401.

\end{acknowledgments}

\end{CJK*}
\end{document}